\documentclass[aps,pre,reprint,groupedaddress,longbibliography,floatfix]{revtex4-1}

\usepackage{amsmath}
\usepackage{amssymb}
\usepackage{graphicx,psfrag}
\usepackage{verbatim}
\usepackage[usenames]{color}
\usepackage{algorithm}
\usepackage{algpseudocode}

\begin{document}
\definecolor{com}{rgb}{0.6,0.1,0.4}

\title{Mirror states enable lower viscosity lattice gases}


\author{Noah Seekins}
\email{noah.seekins@ndsu.edu}
\author{Alexander J. Wagner}
\email[]{alexander.wagner@ndsu.edu}
\homepage[]{www.ndsu.edu/pubweb/$\sim$carswagn}
\affiliation{Department of Physics, North Dakota State University, Fargo, North Dakota 58108, USA}


\date{\today}

\begin{abstract}
    We developed a method for significantly lowering the viscosity achievable for a hydrodynamic lattice gas method. The key advance is the derivation of a mirror state that allows for a reduction of viscosity by more than an order of magnitude over existing lattice gas methods.
\end{abstract}


\maketitle

\noindent\textit{Introduction}
Lattice gas methods are uniquely suited to simulating fluctuating systems due to retaining the fluctuations that naturally arise from the discreteness of matter~\cite{einstein1906theorie}. Certain systems, such as Brownian motion of colloids and reactive mixtures, depend heavily on microscopic hydrodynamic fluctuations to recover the correct macroscopic behavior~\cite{ladd1988application,Bhattacharjee2015Fluctuating}. Previously, viscosities achievable by lattice gas methods were limited, and here we present a method to overcome this.

The minimum achievable viscosities for Boolean lattice gases could be reduced by nearly a factor of 10 by carefully tuning the collision rules and lattice structure~\cite{Henon1987Viscosity,frisch1987lattice}. Yepez showed that for the Frisch-Hasslacher-Pomeau lattice gas (FHP)~\cite{Frisch1986Lattice}, a method that is defined on a hexagonal lattice grid with velocity vectors connecting each lattice site to its six nearest neighbors, the minimum achievable viscosity was around 0.4 in lattice units~\cite{Yepez1999Classical}. The FHP-III model, which had a single rest particle state and allowed for all possible associated collision rules, was able to achieve a minimum viscosity of 0.12 in lattice units~\cite{d'Humieres1987Numerical}.

The FHP model was not extensible to three dimensions due to the lack of a 3D isotropic analogue of the FHP hexagonal lattice. A 4D model using a face-centered-hypercubic (FCHC) lattice was projected down to three dimensions to allow 3D systems to be simulated by lattice gases~\cite{FCHC}. The more complex collision rules of these FCHC lattice gases allowed for an even lower minimum viscosity of approximately 0.048 to be achieved~\cite{Henon1987Isometric,Rivet1988Simulating}. This is believed to be close to the lowest achievable viscosity for Boolean lattice gases.

In related lattice Boltzmann methods further progress was made in lowering viscosities. The Boltzmann approximation was originally used as a theoretical tool to extract the hydrodynamic equations governing lattice gases and with that the viscosity of these lattice gas systems~\cite{frisch1987lattice}. This lattice Boltzmann equation could be simulated directly, which eliminated fluctuations but did not affect the viscosity~\cite{mcnamara1988use}. By simplifying the collision operator to use the Bahatnagar-Gross-Krook (BGK) form, the link to lattice gases was broken~\cite{qian1992lattice}. This new lattice BGK collision operator allowed for viscosities to be lowered arbitrarily (at the cost of losing unconditional numerical stability). 

The BGK collision operator relaxes the distribution function towards equilibrium with an inverse relaxation time $\omega_\mathrm{LB}$, \textit{i.e.} at $\omega_\mathrm{LB}=1$ the collisions will result in a local equilibrium distribution~\cite{qian1992lattice}. The theoretical viscosity is given by,
\begin{equation}
    \nu_{BGK}=\theta\left(\frac{1}{\omega_\mathrm{LB}}-\frac{1}{2}\right),
    \label{eqn: BGKVisc}
\end{equation}
where $\theta=c_s^2\approx1/3$ is the lattice speed of sound. Setting $\omega_\mathrm{LB}\in(1,2]$, \textit{i.e.} relaxing past the equilibrium distribution then leads to a lowered viscosity. This is often referred to as "over relaxation". Overrelaxation theoretically allows for a viscosity of zero at $\omega_\mathrm{LB}=2$, though the lattice BGK method is prone to instabilities for low viscosity systems, and as the viscosity approaches zero the resulting turbulent structures are not resolved by the grid, making the results somewhat unphysical in this limit.

B$\mathrm{\ddot{o}}$sch and Karlin showed that overrelaxation is not a continuous process, and cannot be derived from kinetic theory~\cite{bosch2013exact}. Instead, they showed that replacing the local distribution with a mirror state and underrelaxing ($\omega_\mathrm{LB}\in[0,1)$) from there is equivalent to standard overrelaxation. This mirror state was used by Strand and Wagner to implement overrelaxation in a diffusive sampling lattice gas model based on our prior work~\cite{strand2022overrelaxation, seekins2022integer}.

Here we develop appropriate mirror states to implement overrelaxation in hydrodynamic lattice gas systems. Multiple integer lattice gas algorithms exist, such as the multiparticle lattice gas by Chopard \textit{et al.}~\cite{chopard1998multiparticle}, the collision based integer lattice gas by Blommel and Wagner~\cite{blommel2018integer}, and our sampling lattice gas algorithm~\cite{Seekins2025Integer}. Since the Chopard \textit{et al.} algorithm actually uses a lattice Boltzmann collision as part of the alogorithm it should be possible to use it in combination with overrelaxation, but the minimum achievable viscosity in this algorithm is untested. We are not considering this here since this method utilizes the polynomial lattice BGK equilibrium distribution to sample their local equilibrium ensemble, which Blommel and Wagner showed does not recover the correct local equilibrium behavior~\cite{blommel2018integer}.

In this letter we show that it is possible to lower viscosities using overrelaxation in lattice gases. For simplicity we use our one-dimensional sampling lattice gas algorithm as a proof of concept~\cite{Seekins2025Integer}. In the process we show that correlations retained in lattice gases can lead to non-hydrodynamic behavior and that these correlations become more pronounced for lower viscosities.

\noindent\textit{Sampling Lattice Gas\label{sec: Sampling Algorithm}}
The hydrodynamic sampling lattice gas algorithm \cite{Seekins2025Integer} defines integer lattice occupation numbers $n_i(x,t)$ where $x$ is a lattice cell, $t$ is the iteration number and the $i$ is associated with lattice displacement $v_i$. The $n_i$ are the number of particles moving from cell $x-v_i$ at time $t-1$ to cell $x$ at time $t$ where the lattice spacing and time step are chosen to be one for convenience.
 These occupation numbers then evolve according to the lattice gas evolution equation,
\begin{equation}
    n_i(x+v_i, t+1)=n_i(x, t)+\Xi_i[\{n_j\}_l(x,t)],
    \label{eqn: lgEvo}
\end{equation}

We restrict ourselves to the simplest possible hydrodynamic lattice gas system, a one-dimensional, three-velocity (D1Q3) system. For the hydrodynamic sampling lattice gas, the collision operator is most efficiently defined in moment representation. The three hydrodynamic moments for a D1Q3 system consist of the two conserved moments: the local mass $N$ and the local momentum $J$,
\begin{align}
    N&=n_{1}+n_0+n_{-1},
    \label{eqn: N D1Q3}\\
    J&=n_{1}-n_{-1},
    \label{eqn: J D1Q3}
\end{align}
and a non-conserved moment $\pi$,
\begin{equation}
    \pi=n_{1}+n_{-1}.
    \label{eqn: Pi D1Q3}
\end{equation}
In our notation, the velocity subscripts are given by their numerical value (i.e. $v_i=i$). These moments therefore define the local particle state $\{n_j\}_l$, as each occupation number $n_i$ can be calculated using the three moments,
\begin{align}
    n_1&=\frac{\pi+J}{2},
    \label{eqn: n1 Moment Def}\\
    n_0&=N-\pi,
    \label{eqn: n0 Moment Def}\\
    n_{-1}&=\frac{\pi-J}{2}.
    \label{eqn: nm1 Moment Def}
\end{align}
Conceptually we randomly select particles with probability $\omega$ to be among the particles participating in a collision. The selected particles form occupation numbers $n^\omega_i$. In practice we can sample these occupation numbers from a binomial distribution with probability
\begin{equation} 
    P(n^\omega_i) =\left( \begin{array}{c}
        n_i\\
        n^\omega_i
    \end{array} \right) \omega^{n^\omega_i} (1-\omega)^{n_i-n^\omega_i}.
\end{equation}
The total number of un-collided particles $n_i^{uncol}$ is then given by,
\begin{equation}
    n_i^\mathrm{uncol}=n_i-n_i^\omega.
\end{equation}

As $N$ and $J$ are conserved, $\pi$ is the only moment that is allowed to change over the course of a collision.  Thus, we can write our collision operator in moment representation as,
\begin{equation}
    \Xi_\pi=\hat{\pi}^\omega-\pi^\omega+\pi^\mathrm{uncol}.
    \label{eqn: Col Op Pi}
\end{equation}
We define $\pi^\omega$, $N^\omega$, $J^\omega$ as the moments of the $n_i^\omega$ and $\pi^\mathrm{uncol}$ the second moment of the $n_i^\mathrm{uncol}$ values, each set being used with Eq. (\ref{eqn: N D1Q3}-\ref{eqn: Pi D1Q3}) to calculate these moments. We then pick the post-collision $\hat{\pi}^\omega$ with the local equilibrium probability $P_0(\hat{\pi}^\omega; N^\omega, J^\omega)$.  Deriving the local equilibrium ensemble was a key result of our previous paper, and it is given recursively by~\cite{Seekins2025Integer},
\begin{align}
    &P^0(\pi^\omega+2;N^\omega,J^\omega)=\label{eqn: Pi Dist Recursive}\\
    &\frac{(N^\omega-\pi^\omega)(N^\omega-\pi^\omega-1)}{4((\pi^\omega+2)^2+(J^\omega)^2)}P^0(\pi^\omega;N^\omega,J^\omega).\nonumber
\end{align}
The above equation has defined values starting at $\pi^\omega=|J^\omega|$, and increasing by two while $\pi^\omega\leq N^\omega$. All other values of $\pi^\omega$ have a probability of zero. Eqs. (\ref{eqn: n1 Moment Def}--\ref{eqn: nm1 Moment Def}) then define the set of post-collision $n_i^\omega$, with their respective moments given by $N^\omega$, $J^\omega$, and $\hat{\pi}^\omega$. This set of $n_i^\omega$ values, added to the $n_i^\mathrm{uncol}$ values, gives the full post-collision occupation state.

In our previous paper \cite{Seekins2025Integer} we showed that in many cases the viscosity of the lattice gas is well approximated by eqn. (\ref{eqn: BGKVisc}) with $\omega_\mathrm{LB}$ replaced with the collision probability $\omega$. Since probabilities are required to be between zero and one, there is no choice of $\omega$ that will lead to overrelaxation. However, following Karlin \cite{bosch2013exact}, we can define a mirror state which is then combined with a normal collision step (with the probability $\omega\in [0,1]$) to obtain overrelaxation in our lattice gas. Following the expectation of the lattice Boltzmann result we would expect to get an effective viscosity given by eqn. (\ref{eqn: BGKVisc}) with
\begin{equation}
    \omega_\mathrm{eff}=\left\{\begin{array}{ll}
    \omega & \mathrm{without\; applying\; mirror\; state}\\
    2-\omega & \mathrm{when\; mirror\; state\; is\; applied}
    \end{array}\right.
    \label{eqn: omegaeff}
\end{equation}
which would be the corresponding lattice Boltzmann result, when formulated using a mirror state. The effectiveness of this assumption is tested below.

\noindent\textit{Mirror State}
For a diffusive lattice gas Strand and Wagner showed that a suitable mirror state consists of simply flipping the velocites~\cite{strand2022overrelaxation}. In the hydrodynamic case, we must instead create a mirror state around the local equilibrium $\pi$ ensemble since the diffusive mirror state doesn't conserve momentum. 

For a state to qualify as a mirror state over the local equilibrium ensemble, given an initial state $\{n_j\}_l$, it must fulfill the following conditions: firstly, the mirror state {$\{n_j\}^m_l=M(\{n_j\}_l)$ must have the same $N$ and $J$ values as the original state. The mirror operation shouldn't move a system out of equilibrium. Given the discreteness of $\pi$ and the non-symmetric nature of the local equilibrium $\pi$ ensemble there cannot be a one-to-one correspondence between $\pi$ and $\pi^m=M(\pi)$ in general, which means that the mirror operator has to be probabilistic. We can therefore write the condition that the local equilibrium $\pi$ ensemble needs to be invariant under the mirror operation as
\begin{equation}
    P^0(\pi;N,J)=\sum_{\pi'}\left\langle\delta_{\pi,M(\pi')}P^0(\pi';N,J)\right\rangle^M,
\end{equation}
where $\langle...\rangle^M$ indicates an average over the outcomes over the probabilistic mirror operator $M(\pi')$, and $\delta_{\pi,M(\pi')}$ is the Kronecker delta.

Lastly the mirror operation should be as invertible as possible such that $\langle \{M[M(N,J,\pi)]-\pi\}^2 \rangle$, the average distance from the initial $\pi$ value of the $\pi$ value of a twice-mirrored state, is minimized. 

These conditions imply a construction for the mirror population that we believe to be unique. This construction starts with the equilibrium $\pi$ distribution corresponding to the local $N$ and $J$ values. We construct the forward and backward cumulative distributions corresponding to the local equilibrium ensemble, which are explicitly given by,
\begin{align}
    &C(\pi; N,J)=\sum_{\pi'=0}^{\pi} P_0(\pi'; N,J),\label{eqn: Cumulative}\\
    &C_{-1}(\pi; N,J)=\sum_{\pi'=\pi}^{N} P_0(\pi';N,J), \label{eqn: Anti-Cumulative}
\end{align}
where $\pi'$ is an index that enumerates the possible values of $\pi$.

\begin{figure}
    \centering
    \includegraphics[width=0.35\columnwidth]{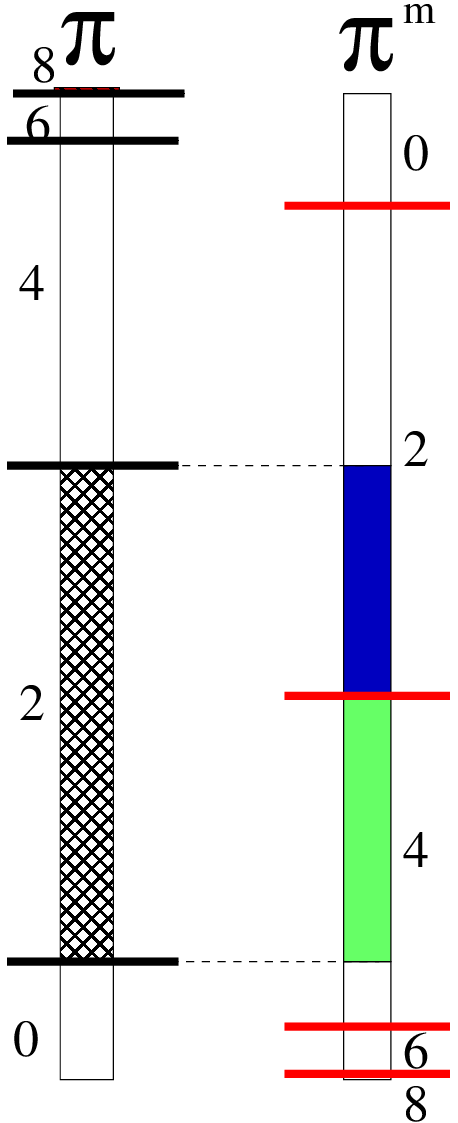}
    \caption{An illustration of the flipping operation for $N=9$, $J=0$, and an initial $\pi$ value of $2$. The forward ($\pi$) and backward ($\pi^m$) cumulative distributions are given by} Eqs. (\ref{eqn: Cumulative}) and (\ref{eqn: Anti-Cumulative}) respectively. See text for further details.
    \label{fig: Visual Flipping Example}
\end{figure}
Graphically the mirror operation is presented in Fig.~\ref{fig: Visual Flipping Example}. A value of $\pi$ corresponds to the shaded region on the left, and this area is mapped onto the backwards cumulative distribution on the right. We then pick a uniformly distributed random point in the shaded area on the left and find a corresponding $\pi^m$ value in the backward cumulative distribution. The chosen $\pi^m$ value is then the selected mirror state value for the initial $\pi$ value. Mathematically this can be expressed as,
\begin{align}
    &C_{-1}(\pi^m;N,J)\leq\\
    &C(\pi-2;N,J)+rP_0(\pi;N,J)< C_{-1}(\pi^m+2;N,J),\nonumber
\end{align}
where $r$ is a uniformly distributed random number between $0$ and $1$. In Fig. \ref{fig: Visual Flipping Example}, the shaded region on the left (corresponding to an initial $\pi$ value of two) can either be mapped to the darker or lighter (blue and green online) shaded regions (corresponding to $\pi^m$ values of two and four respectively) depending on the selected value of $r$.

The fact that this transformation is not one-to-one means that when the mirror operator is applied repeatedly it will cause the distribution of $\pi$ values to spread out, eventually recovering the whole local equilibrium $\pi$ ensemble. The rate of this effect is also dependent on the average density of the observed system, spreading out more slowly for higher density systems. Though easy to observe in a system without streaming, the impact of this spreading in a system with streaming is outside the scope of this letter, and thus will be left to future work.

\noindent\textit{Simulation Results\label{sec: Results}}
To measure the viscosity of our system, we consider the decaying sound wave used in our previous paper for the same purpose. The system is initialized close to global equilibrium, with a small sinusoidal perturbation. The amplitude of this perturbation is on the order of 1\% of the average density in equilibrium, $\bar{N}^{eq}$. This amplitude then oscillates, with maxima that relax towards equilibrium exponentially, with a decay rate given by~\cite{luizThesis}
\begin{equation}
    \lambda=\left(\frac{2\tilde{\pi}}{L}\right)^2\nu.
    \label{eqn: nuAndlambda}
\end{equation}
This decay rate depends on the kinematic viscosity of the system $\nu$ and the system size $L$, and $\tilde\pi$ is the mathematical constant. Thus, to recover the kinematic viscosity, we extract the amplitude of our sine wave using a process developed by Blommel and Wagner~\cite{blommel2018integer}. We then extract the maxima of this amplitude to better visualize the decay.

In our prior paper, we noted that there was some difference between the viscosity at early times ($t<300$ timesteps in that paper) and late times ($400<700$ timesteps in that paper)~\cite{Seekins2025Integer}, especially for low densities. This is a deviation from the analytical solution, which assumes a constant decay rate. However, we find a good fit if we assume the viscosity of this system is a function of time. Along with a numerical analysis of the collision operator done in that paper, the deviation implies that the lattice gas is capturing correlation dependent phenomena that break the assumption underlying the Boltzmann approximation~\cite{Huang1987Statistical}. The time dependence of the viscosity implies that correlations can become important in these lattice gas systems. In the supplemental material we show that two ensembles with the same Boltzmann average can evolve differently if different correlations are present, and that such correlations develop naturally in the case of a decaying sin wave. 

\begin{figure}
    \centering
    \includegraphics[width=\columnwidth]{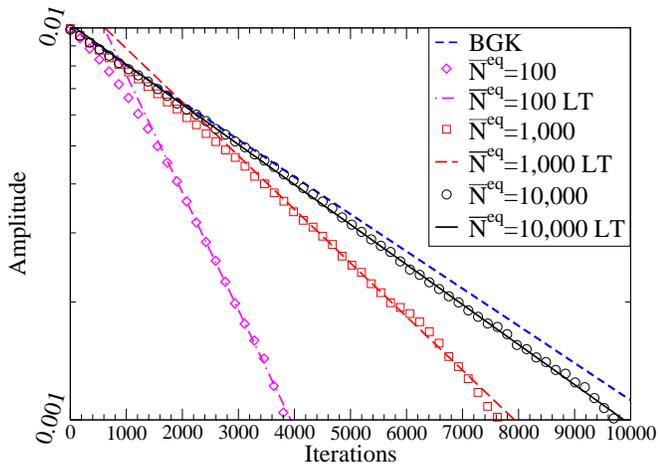}
    \caption{Extracted maxima of decaying sound waves with initial average densities of $\bar{N}^{eq}$=100, 1,000, and 10,000 and a system size of 100 lattice sites, shown on a logarithmic scale. All systems were initialized identically to those in our prior paper, and run for 30,000 timesteps with an $\omega_\mathrm{eff}$ value of 1.5, and then averaged over 2,500,000/$\bar{N}^{eq}$ random seeds. The theoretical decay rate of a BGK system calculated using Eqs. (\ref{eqn: BGKVisc}) and (\ref{eqn: nuAndlambda}) is also shown, along with late time fittings (LT) for each density generated via xmgrace non-linear curve fitting. See text for additional details about the late time fittings.}
    \label{fig: DecayRate}
\end{figure}
Fig. \ref{fig: DecayRate} shows an example of how the extraction of the decay rate was performed, as well as the relative scaling of this effect for different densities. For $\bar{N}^{eq}$=10,000, there is good agreement with the BGK for approximately the first 5,000 timesteps, however after that the system deviates, and the late time fitting agrees better. This is also true for $\bar{N}^{eq}$=1,000 and $\bar{N}^{eq}$=100, though their divergences from the BGK happen earlier. Their late time fittings also show far more clearly their difference from the early time data. 

The late time fittings themselves are visually approximated at the point at which the amplitude decay becomes linear on the log plot, as we observed the eventual late time decay rate to be relatively constant. Out simulations show the exponential decay for about an order of magnitude before losing coherent signal to noise, which sets the scale of Fig. \ref{fig: DecayRate}. 

We found no explicit discussion of the importance of correlations to viscosity values in the literature of Boolean lattice gas systems. However, in work done by Yepez, the decay graphs for the decaying sound wave had fittings that failed to match the visible decay at later times~\cite[Fig. 4.12]{Yepez1999Classical}. So the importance of correlations in lattice gas systems leading to dynamics beyond the Boltzmann limit is likely not limited to our lattice gas systems, but may exist more widely.

\begin{figure}
    \centering
    \includegraphics[width=\columnwidth]{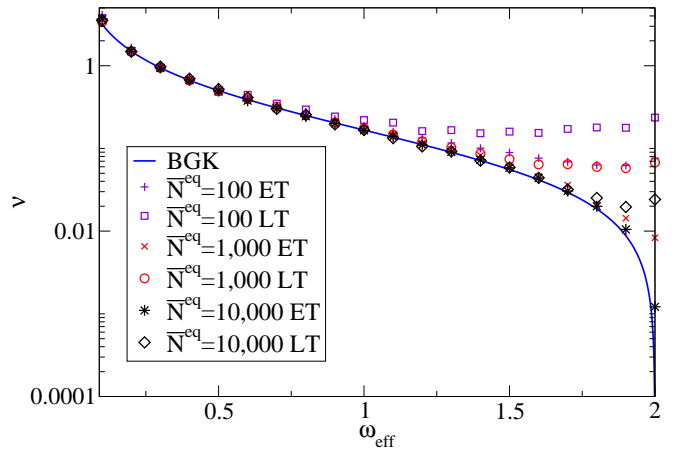}
    \caption{The viscosity recovered for early time (ET - $t<500$) and late time (LT - found via the method used in Fig. \ref{fig: DecayRate}) systems, shown on a logarithmic scale. These systems were initialized and run identically to those in Fig. \ref{fig: DecayRate} for $\omega_\mathrm{eff}$ ranging from 0.1 to 2. The error in these viscosities, calculated by simulating the same system using multiple sets of random seeds, is on the order of the symbol size. Eq. (\ref{eqn: BGKVisc}) is also shown.}
    \label{fig: EarlyLate}
\end{figure}
Fig. \ref{fig: EarlyLate} shows the recovered early and late time viscosity values over while varying $\omega_\mathrm{eff}$ for $\bar{N}^{eq}$=100, 1,000, and 10,000. These results are compared to the viscosity given by the BGK model from Eq. (\ref{eqn: BGKVisc}). The BGK and measured viscosities show good agreement for early time and high density systems. In our prior paper, we observed that the viscosity of a system will increase for lower average densities. We examined this effect for several densities at $\omega_\mathrm{eff}=1.0$, \textit{i.e.} in the absence of a mirroring operation. Fig. \ref{fig: EarlyLate} shows that this phenomenon continues continuously into the overrelaxation regime. We observe that each density has an associated minimum viscosity. Both the early and late time viscosity values are shown here, and we see that the early time results in the absence of correlation agree notably better with the BGK values, where as the late time viscosities are significantly higher, although both early and late time viscosities show a minimum value.

As mentioned earlier, the specially tuned FCHC model was able to lower viscosities in Boolean lattice gases to 0.048 in lattice units. The late time viscosity of a system with a density of 10,000 particles improves on this by a factor of about 2.5, whereas the early time viscosity shows a reduction by a factor of about 41. This early time viscosity is likely the value reported in the literature because most of those papers did not consider an sufficient oscillations to observe late time phenomena~\cite{Henon1987Viscosity, frisch1987lattice, d'Humieres1987Numerical, Henon1987Isometric, Rivet1988Simulating}. As the minimum possible viscosity decreases with density, we believe our algorithm can achieve even lower viscosities. However, we are currently limited to simulating a maximum of about 20,000 particles reliably, as the memory use of the lookup tables used to efficiently calculate the mirror state for a given set of $N$ and $J$ values scales with $O(N^2)$. Along with this, at larger values of $N$ the current algorithm struggles to numerically resolve the tails of the distribution.

Remarkably at $\omega_\mathrm{eff}=2$, i.e. with only the mirror state transformation performed, the viscosity increases very slightly over $\omega_\mathrm{eff}=0.9$. This curious behavior is somewhat counter-intuitive, but is outside the scope of this letter.


\noindent\textit{Outlook}
We have presented a mirror state that allows for overrelaxation to be implemented in a 1D hydrodynamic lattice gas. We showed that this mirror state allows us to achieve lower viscosities compared to those that could be achieved in Boolean lattice gas systems by over an order of magnitude.

Some algorithmic advances are expected to lower these viscosities even further by allowing for larger numbers of particles per cell.
Preliminary findings have shown that the local equilibrium $\pi$ ensemble is well approximated by a normal distribution in the limit of high $N$. Sampling our local equilibrium ensemble from a normal distribution, as was done by Chopard \textit{et al.}~\cite{chopard1998multiparticle}, would improve the memory scaling with respect to density to $O(1)$ instead of $O(N^2)$. Additionally special care will need to be taken in this limit for systems far out of equilibrium that show states in the tails of the equilibrium distribution where the tiny probabilities are difficult to represent as floating point numbers.

An alternate approach to decrease the viscosity consists of avoiding the application of mirror states moving the system closer to equilibrium. A mirror operator could be engineered to prevent the drift by intentionally moving the system out of local equilibrium by an amount that on average opposes the effective numerical drift. However, this process would result in global equilibrium systems being moved out of equilibrium, and thus has not been considered by us yet.

The current implementation is only one-dimensional, but the original Blommel algorithm worked in any dimension~\cite{blommel2018integer}. We are currently working on extension of the sampling algorithm to higher dimensions. Implementing the mirror state in higher dimensions leads to interesting degrees of freedom. In the diffusive overrelaxation algorithm~\cite{strand2022overrelaxation} only the $J$ mode was altered and the $\pi$ mode was left invariant. Here only the $\pi$ mode is altered because, in a momentum conserving D1Q3 model, this is the only mode that can undergo a mirror operation. In higher dimensions additional moments controlling both the bulk and shear viscosities as well as several non-hydrodynamic modes can participate in the mirror operation, leading to a more complex choice of possible mirror states.

Of great theoretical interest is our observation in this letter that the viscosity evolves in time. This indicates that the development of significant correlations leads to physics beyond the Boltzmann limit. Characterizing these correlations and understanding how they influence the viscosity is an important subject that we hope to address in the near future.

Just as important is the question if this non-Boltzmann behavior is restricted to lattice gases, or if it is also observed in molecular systems, as simulated with Molecular Dynamics. A mapping procedure between Molecular Dynamics and lattice gases exists \cite{parsa2017lattice}. Using this approach Pachelieva and Wagner showed that overrelaxation can be recovered from a continuous molecular dynamics simulation coarse-grained onto a lattice gas system~\cite{pachalieva2021connecting}. Establishing that the non-Boltzmann behavior observed here is also present in systems simulated by Molecular dynamics would add significant weight to these observations.

\appendix


\bibliographystyle{apsrev4-2}
\bibliography{Bibs/AW,Bibs/MCLG,Bibs/IntegerLG,Bibs/LB,Bibs/SODShockTube,Bibs/HiReyn}

\end{document}


\definecolor{com}{rgb}{0.6,0.1,0.4}

\title{Supplemental material to: Mirror states enable lower viscosity lattice gases}


\author{Noah Seekins}
\email{noah.seekins@ndsu.edu}
\author{Alexander J. Wagner}
\email[]{alexander.wagner@ndsu.edu}
\homepage[]{www.ndsu.edu/pubweb/$\sim$carswagn}
\affiliation{Department of Physics, North Dakota State University, Fargo, North Dakota 58108, USA}


\date{\today}

\begin{abstract}
    We demonstrate here that the decaying isothermal sound waves considered in our letter to determine the viscosity can show non-Boltzmann behavior, \textit{i.e.} there is no unique Boltzmann limit for our lattice gas. Instead correlations in the lattice gas can build up and lead to significant changes in the viscosity, in some extreme cases increasing the viscosity by a factor of 10.
\end{abstract}

\keywords{Lattice Gas, Lattice Boltzmann, Monte Carlo, Fluctuatons, Overrelaxation}

\maketitle


In the main text we made the strong claim that we observed non-Boltzmann behavior in the simulations of a decaying sine wave. In practice what this means is that two simulations having identical Boltzmann averaged $f_i=\langle n_i \rangle$ does not guarantee identical future evolution. We showed previously that the collision operator can, in principle, be altered by highly correlated $n_i$, even though such highly correlated configurations don't often evolve in practical simulations, making this possibility mostly academic~\cite{Seekins2025Integer}. Instead the Boltzmann average of the collision operator is often very well approximated by a BGK collision operator.

When we examined the decay of a stationary sound wave in our previous paper \cite{Seekins2025Integer} we observed two non-trivial phenomena. Firstly the decay rate did depend on density. Some dependence is clearly expected, since for very low densities it will become harder to find collision partners for particles, but the observed effect was more pronounced than would be expected. Secondly we found that the decay of the maxima of the sound wave did not follow a simple exponential decay, but instead we observed that the decay rate appeared to depend on time. In the current letter we showed that this is well modeled by a system with two decay rates, transitioning from an early to a late time value. The apparent change of the value of the viscosity is significant: if the viscosity really changes as a function of time (rather than as a function of the hydrodynamic fields), then clearly the hydrodynamic equations are not sufficient to describe the system's behavior.

\begin{figure}[h]
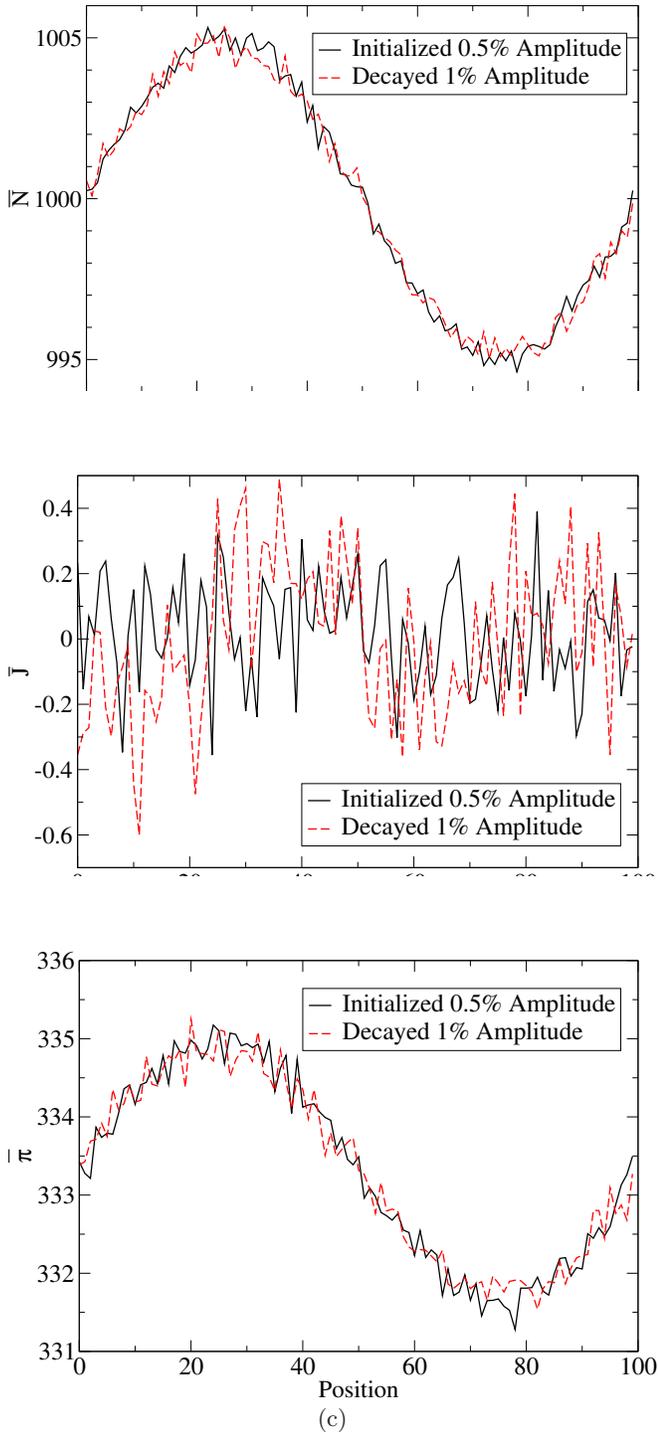

    \centering
    \includegraphics[width=\columnwidth]{Graphs/NHalfComp.eps}\\(a)\\
    \includegraphics[width=\columnwidth]{Graphs/JHalfComp.eps}\\(b)\\
    \includegraphics[width=\columnwidth]{Graphs/PiHalfComp.eps}\\(c)
    \caption{The $\bar{N}$ (a), $\bar{J}$ (b), and $\bar{\pi}$ (c) hydrodynamic fields for two lattice gas systems. The first lattice gas system (solid line - black online) was initialized identically to those discussed in Fig. 2 of the main text with an average density of $\bar{N}^{\mathrm{eq}}=1,000$, a system size of $L=100$, and an initial amplitude of $1\%$ of its average density. It was then decayed until it reached $50\%$ of its initial amplitude. The second lattice gas system (dashed line - red online) was initialized with an initial amplitude of $0.5\%$ of its average density, but otherwise was initialized identically to the first, and was not run, but was averaged over 5,000 random seeds.}
    \label{fig: NJPiHalfComp}
\end{figure}

The fact that the extracted viscosity is not constant could of course have other explanations. In particular the shape of the sound wave could change, showing some inertial effects, rendering the analytical solution invalid or the non-equilibrium values of the $\pi$-mode could evolve different in time. This is why we felt that it was necessary to substantiate our claim of non-Boltmzmann behavior in this supplemental material. 

To substantiate our claim we performed the following simulation: first we observed the decay of a sine wave which showed different early and late time behavior. We evolved this configuration until we reached an oscillation peak at which the amplitude was about half its original value. We then initialized a second system with independently Poisson distributed $n_i$ corresponding to the analytical prediction of the moments $N$, $J$ and $\pi$. 

We verified that the resulting Boltzmann averages over the hydrodynamic moments of these two sine waves were identical, up to remaining noise. This is shown in Fig. \ref{fig: NJPiHalfComp} for $\omega_{\mathrm{eff}}=1.5$, the same value that we used in the main text for our illustrations. We see that the three average moments agree up to remaining noise, with no notable deviation. This shows that the freshly initialized ensemble and the ensembles that have already decayed to half the amplitude have the same Boltzmann average. Now, if a unique Boltzmann average existed, these two ensembles should evolve in the same manner. 

\begin{figure}
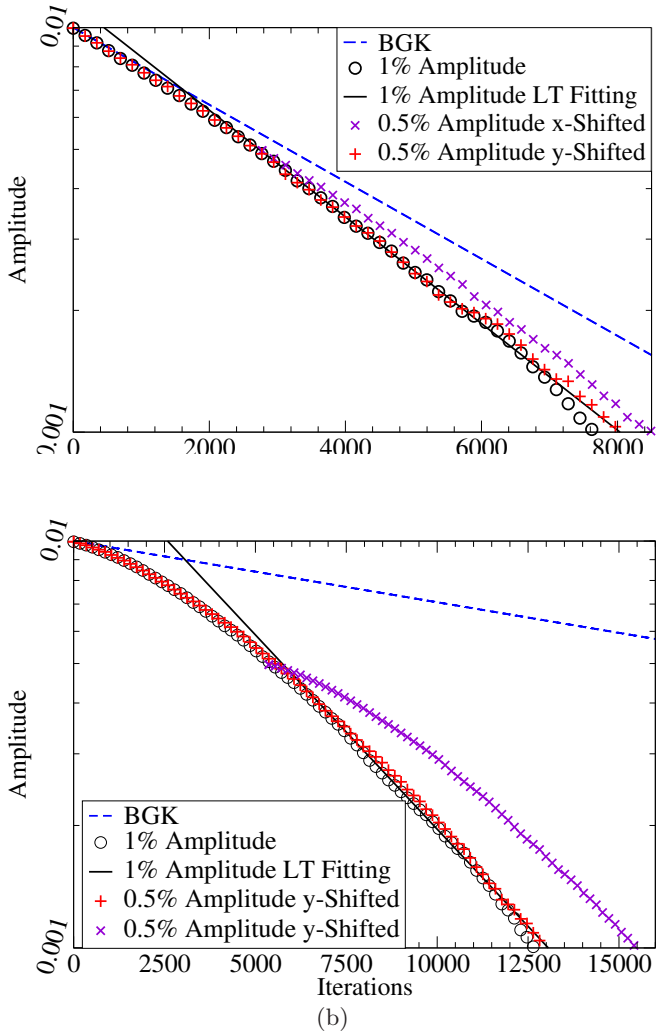

    \centering
    \includegraphics[width=\columnwidth]{Graphs/HalfDecayTest.eps}\\(a)\\
    \includegraphics[width=\columnwidth]{Graphs/HalfDecayComp1p9.eps}\\(b)
    \caption{Extracted maxima of two decaying sound waves with an average initial density of $\bar{N}^{\mathrm{eq}}=1,000$ for $\omega_{\mathrm{eff}}=1.5$ (a) and $\omega_{\mathrm{eff}}=1.9$ (b). These lattice gas systems were initialized identically to those in Fig.~\ref{fig: NJPiHalfComp}, and were run identically to those in Fig. 2 of the main text, aside from the second system being averaged over 5,000 random seeds to account for the initially halved amplitude. The amplitude data for the $0.5\%$ initial amplitude system has been scaled to match that of the $1\%$ amplitude system for both x ((a) shifted by +2775 iterations in time and (b) shifted by +5370 iterations) and y (initial amplitude maxima for both (a) and (b) scaled to match that of their respective $1\%$ amplitude system) for the purpose of comparing their decay rate. The theoretical decay of a BGK system is also shown.}
    \label{fig: AmpDecayComp}
\end{figure}

What we observe, however, is quite different. In Figure \ref{fig: AmpDecayComp} (a) we see that the x symbols showing the evolution of the freshly initialized system does evolve differently from the already decayed system shown as o symbols. However, when you rescale the freshly initialized system by a factor of two and match the zero points in time, we observe that the evolution, including early and late time is statistically identical to the first simulation. This shows that there is early and late time decay independent of the initial amplitude. While the exact nature of the correlations that cause this behavior is unexplained so far, this convincingly shows that these integer lattice gases show non-Boltzmann behavior. 

This effect becomes even more striking when one moves to lower viscosities, as we showed in our letter. So for illustration we repeated the procedure for $\omega_\mathrm{eff}=1.9$. The result of these simulations are shown in Figure \ref{fig: AmpDecayComp}(b). The deviation from Boltzmann behavior is even more pronounced here. 

As pointed out in the main text we believe this to be a potentially important discovery. However, we would like to understand better what the nature of the underlying correlations are that build up in these simulations and whether such correlations are also common in other discrete systems like Molecular Dynamics simulations.

\appendix


\bibliographystyle{apsrev4-2}
\bibliography{Bibs/AW,Bibs/MCLG,Bibs/IntegerLG,Bibs/LB,Bibs/SODShockTube,Bibs/HiReyn}